\documentclass[onecolumn,preprintnumbers,amsmath,amssymb,nofootinbib]{revtex4}
\usepackage[english]{babel}
\usepackage{graphicx}

\begin{document}
\def\p{\partial}
\def\e{\varepsilon}
\def\f{\varphi}
\def\t{\tilde}
\def\ds{\displaystyle}


\title{Dissipationless Disk Accretion}

\author{\firstname{S.~V.}~\surname{Bogovalov,}}
\author{\firstname{S.~R.}~\surname{Kelner}}
\affiliation{%
Moscow Engineering Physics Institute (State university), Moscow, Russia
}%

\begin{abstract}
We consider disk accretion resulting purely from the loss of angular
momentum due to the outflow of plasma from a magnetized disk.
In this limiting case, the dissipation due to the viscosity
and finite electrical conductivity of the plasma can be neglected.
We have obtained self-consistent, self-similar solutions for dissipationless
disk accretion. Such accretion may result in the formation of
objects whose bolometric luminosities are lower than the flux of kinetic
energy in the ejected material.

\end{abstract}
\maketitle

\section{INTRODUCTION}
Disk-accretion mechanisms can provide clues
about a number of processes occurring when matter
is accreted onto a gravitating center. It is thus no
wonder that such mechanisms have been intensely
studied for more than 30 years. The detection of
directed ejections from both the cores of active
galaxies [1] and Galactic and intergalactic sources,
such as SS433 [2], young stellar objects [3], and
Galactic super luminal sources [4], has revealed an
unambiguous connection between jets and accretion
disks. The presence of a directed ejection in a source
is essentially always associated with evidence for the
presence of an accretion disk [3]. One exception is the
jet like structures in plerions [5], whose nature is obviously
different [6]. Thus, the collected observational
data indicate that disk accretion is frequently (or
even always) accompanied by the narrowly collimated
ejection of a substantial fraction of the accreted
matter from the source.
This picture gives rise to a number of problems.
What is the mechanism that leads to some fraction
of the matter falling on to the gravitating center
being ejected from the source? How is this matter
collimated? What makes the collimated ejection stable
enough to propagate to great distances from the
source? What accelerates the charged particles that
generate the observed radiation in the jets? This list
of questions is far from complete. Here, we consider
just one of them: what mechanism ejects some of the
matter falling on to a gravitating center in the course
of disk accretion?
To fall onto a gravitating center, matter with nonzero
angular momentum must somehow lose angular
momentum. This angular momentum must be carried
away by matter, so that disk accretion should
inevitably be accompanied by an outflow of matter;
the only problem is to determine its rate and direction.
In the classical theory of . accretion disks [7],
angular momentum of the accreted matter is lost
via viscous, hydrodynamical, turbulent stresses. The
classical theory does not include a magnetic field. In
[8], however, the magnetic field was used to provide
high viscosity. This study also considered the possibility
of a matter outflow along the axis of rotation of
the accretion disk in a hypercritical regime, when the
radiation pressure exceeds the gravitation. A quantitative
theory of such an outflow is given in [9], which
shows that the outflow already begins in a state with
subcritical luminosity. This ejection mechanism may
be important in some sources with hyper-Eddington
luminosities; however, it is clear that, in many cases
(e.g., in young stellar objects), the ejection is driven
by another factor, since the radiation pressure in these
objects is too small to support this mechanism. Even
in SS433, the ejection energy cannot be provided
purely by radiation pressure [10].
In standard disk-accretion theory, angular momentum
is lost via viscous stresses due to the differential
rotation of the disk. Radiation-efficient disks
are geometrically thin, with a local temperature that
is substantially lower than the virial temperature. It
is very difficult to provide an appreciable outflow from
the disk under these conditions, since the disk matter
is in a very deep potential well. Recently, accretion
in radiatively inefficient disks (ADAFs) for which the
ion temperature is close to the virial temperature have
been widely discussed [13, 14]. Generally speaking,
the matter in such disks can flow out vigorously,
with the subsequent formation of a jet. However, the
existence of ADAF disks is in doubt [15]. Therefore,
it remains important to search for alternative mechanisms
for the ejection of accreted matter. Ejections
due to the presence of a magnetic field in the disks are
of particular interest.


Blandford and Payne [16] have suggested one of
the most promising mechanisms for jet outflows.
Matter moves in the disk with close to Keplerian
velocities. If a magnetic field penetrates the disk, the
matter is blown off the disk by centrifugal forces if
the inclination of the magnetic-field lines to the disk
is smaller than $60^\circ$. This mechanism is particularly
easy to understand if we imagine the motion of the
plasma along a magnetic line to be like that of a
bead on a rotating wire. The bead moves in the disk
along a circular Keplerian orbit, with the sum of all
the forces being zero. If a field line is inclined to the
equator at an angle smaller than $60^\circ$, any shift of the
bead along the magnetic-field line will result in an
increase in the centrifugal force and a decrease in
the gravitation force. In this case, the position of the
bead on the disk is unstable. In addition to forming
an ejection, this mechanism can explain the loss of
angular momentum, which is carried away from the
disk by the wind.
The influence of the matter outflow on the dynamics
of the accretion disks themselves was considered
in [17--19] in the context of young stellar objects.
It was shown that the loss of angular momentum
due to the wind from the disk can be more efficient
than losses due to turbulent viscosity. This raises the
issue that disk accretion onto a gravitating center
should take into account the fact that some of the
angular momentum is carried away by the disk wind.
Ferreira [21, 22] attempted to solve this problem.
His formulation is similar to that used previously by
Bisnovatyi-Kogan and Ruzmaikin [20] in one of their
pioneering studies on the theory of disk accretion in a
magnetized plasma. It is assumed that the interstellar
plasma is the source of an almost uniform magnetic
field. In this case, in order for matter to fall onto
the gravitating center, not only must angular momentum
be lost, but matter must also diffuse across
the magnetic field. In addition to a high turbulent
viscosity, this requires a low electrical conductivity
of the plasma, which must be associated with a high
level of turbulence. In other words, dissipation plays a
key role in this case.
We will use another formulation of the problem.
Figure 1a presents the difference between our formulation
and that used by Ferreira et al. [21, 22] (left).
Interstellar matter with some initial angular momentumand
a relatively uniform magnetic field is accreted
onto a gravitating center. The accreted matter diffuses
across the field lines, while the forming wind flows
out along magnetic lines stretched from interstellar
matter.
Figure 1b (right) presents another situation.Matter
is accreted onto the compact object along with
the frozen-in magnetic field, forming the accretion
disk. During the accretion, magnetic lines emerge
from the accretion disk and are stretched. A specific
mechanism for this stretching of magnetic lines
from the disk was proposed and studied numerically
in [23]. It results in the formation of a corona and,
accordingly, a magnetized disk wind. In this case,
the accretion disk itself is a source of the magnetic
field of the outflowing wind, just as the Sun is the
source of the magnetic field in the solar wind. Note
that precisely such a formulation of the problem was
used in the study [16] of the centrifugal mechanism
for the outflow of matter from an accretion disk.
Here, we study not only the outflow itself, as was
done in [16], but also the influence of the plasma outflow
from an accretion disk on the accretion. This is
not only important for the physics of accretion disks,
but is also directly applicable to the interpretation
of the available observational data. Observations of
Galactic superluminal sources [4, 24] are of particular
interest, since they provide conclusive evidence that
plasma ejections substantially affect the accretion
process.

\section{FORMULATION OF THE PROBLEM}
In general, viscous stresses in accretion disks
are undoubtedly important for angular-momentum
transport. Here, we primarily consider the influence
of the wind on the accretion dynamics. Therefore,
we will consider the limiting case when all of the
angular momentum is carried away from the disk by
the wind with no dissipation. The possibility of such
dissipationless accretion can easily be understood
based on the following simple consideration. Let
one of the components of a binary rotate around
a fixed gravitating center (Fig. 2). In the absence
of dissipation and wind, this rotation will continue
indefinitely.
Let us now suppose that a magnetized wind flows
from the rotating object, and that the Alfven radius
is $R_a$. We will assume that the wind is spherically
symmetrical and its velocity is $v_0$. For simplicity, we
will use here the mechanical analogy of beads moving
along wires for the motion of the plasma in a strong
magnetic field. This kind of motion occurs up to the
Alfven radius, beyond which the plasma moves essentially
freely. We assume that there is no intrinsic
rotation of the object, so that its angular momentum
is $M\cdot l_z$, where $M$ is the mass of the object,
and $l_z=\Omega R_0^2$ is the specific angular momentum due to its
orbital motion.
We denote $L$ to be the total angular momentum
contained in a sphere with radius $R_a$ with its center at
the point ${\bf R}_0(t)$ at time $t$. At time $t+\delta t$, the angular
momentum contained in this sphere (with its center
at ${\bf R}_0(t)$), will differ from $L$. The equation for the
variation of the angular momentum has the form
\begin{equation}
\frac{dL}{dt}= -\oint\!\rho\,[{\bf r}_a\,{\bf v}]_z\,v_n \, dS.
\end{equation}
Here $\rho$ is the density of the plasma, $v_n$ is the velocity
normal to the integration surface, the integration is
carried out over the Alfven surface, and the z axis
is directed along the angular-momentum vector. We
can see from Fig. 2 that ${\bf r}_a={\bf R}_0+{\bf R}_a$, the velocity
of the plasma is ${\bf v}=v_0{\bf n}+[{\bf \Omega}\,{\bf r}_a]$,
where ${\bf n}={\bf R}_a/R_a$
is the unit vector in the direction ${\bf R}_a$, and $\bf \Omega$ is the
angular-velocity vector for the rotation about the fixed
center. After integrating over the directions of the
vector $\bf n$, we obtain
\begin{equation}\label{loss}
\frac{dL}{dt}=-\frac23\,\dot M\,\Omega\,(2\,R_0^2+R_a^2)\,.
\end{equation}
To interpret this result, note that $\dot L_M=-\dot M\,\Omega\,R_0^2$
is the variation of the angular momentum due to the
decrease of the mass of the rotating object. It can
be shown that the rate of variation of the angular
momentum of the wind is $\dot L_V=-\dot M\,\Omega\,R_0^2/3$.
This can be determined as the difference between the angular
momenta $L_V (t +\delta t)$ and $L_V(t)$ of the wind particles
in a sphere of radius $R_a$ with its center at the fixed
point $R_0(t)$. The sum $\dot L_M+\dot L_V$ yields the first term
in (2). Therefore, the second term describes the rate
of variation of the specific angular momentum associated
with the orbital motion:
\begin{equation}
\frac{dl_z}{dt}=-\frac23\,\frac{\dot M}{M}\,\Omega\,R_a^2\,.
\end{equation}
Thus, the outflowingwind carries away the angular
momentum of the rotatingobject.

This simple example illustrates an important property
of accretion in the presence of a magnetized wind.
Since the accretion is not accompanied by heating,
the wind carries away not only angular momentum,
but also rotational energy. Dissipationless accretion
can result in the formation of ejections for a relatively
low luminosity of the object. This makes this accretion
mechanism particularly interesting: it can explain
the unusually high efficiency for the transformation of
the gravitational energy of the accreted matter into
the kinetic energy of the jet in the unusual source
SS433. The bolometric luminosity of SS433 is about
$3\cdot 10^{39}$~erg/s [25], while the flux of kinetic energy
from the object is at least twice this value
($>6 \cdot 10^{39}$~erg/s) [12], which is difficult to understand in
classical accretion models [10].
It is reasonable to consider dissipationless accretion
in an ideal magnetohydrodynamical (MHD) approximation.
This leads to the model in the left panel
of Fig. 3: the wind stretches magnetic-field lines from
the accretion disk. The polarity of the lines varies
chaotically, since the total flux of the magnetic field
emerging from either side of the accretion disk is, on
average, zero. At first glance, it would seem impossible
to describe the dynamics of a wind in the presence
of a magnetic field with arbitrarily varying polarity at
the disk surface. However, there is a circumstance
that simplifies the situation radically. As was noted
in [26], the dynamics of an ideal plasma are invariant
with respect to the direction of the magnetic-field
lines. Therefore, if the direction of the field lines varies
so that the polarity of the magnetic field is the same
on each side of the accretion disk, this does not affect
the plasma motion, and we are able to consider a
self-consistent outflow of plasma from an accretion
disk with an azimuthally symmetrical magnetic field,
shown schematically in Fig. 3 (right).
Two features make this approach different from the
well-known formulation of Blandford and Payne [16].
First, the outflowingwi nd must be consistent with the
accretion rate in the disk, whereas, in [16], the disk
and wind parameters were specified independently.
Second, since the plasma moves toward the gravitatingc
enter, there is a nonzero azimuthal electrical
field, $E_{\varphi} \ne 0$. Note that a similar formulation was
given by Contopoulos [27]. However, his study was
not continued, since it was clear that, in this case,
magnetic flux should accumulate at the center. If the
polarity of the field is the same everywhere, it cannot
be annihilated, and the accretion will inevitably be
halted by the pressure of the magnetic field. As we
noted above, this problem is removed because we are
solving for a plasma flow in the presence of a magnetic
field of variable polarity. Therefore, the magnetic flux
does not accumulate at the center, and is always equal
to zero, on average. Since the magnetic-field lines
possess different polarities, they can reconnect, so
that the magnetic-field pressure at the center will not
increase without bound.
Finally, we will assume that the plasma flow in the
wind is self-similar. This assumption is natural if we
are interested in the motion of plasma directly above a
very thin disk away from its edges. Then, the problem
has no parameters with the dimension of length, and
the flow becomes self-similar. This approximation has
been used to describe magnetized winds from accretion
disks in numerous studies, starting with [16].
Vlahakis and Tsinganos [28] present a detailed description
and classification of all possible types of selfsimilarity.
Note also the study [29], which gives selfsimilar
solutions describing nonstationary accretion.

\section{GENERAL RELATIONS}
The dimensionless steady-state equations for an
ideal, cool plasma (with pressure P = 0) can bewritten
(see, for example, [30])
\begin{equation}\label{e1}
\rho({\bf v}\nabla){\bf v}=-\frac12\,\nabla\,{\bf B}^2+({\bf B}\nabla){\bf B}
-\rho\,\frac{g\bf R}{R^3}\,,
\end{equation}
\begin{equation}\label{e2}
{\rm div}(\rho{\bf v})=0\,,
\end{equation}
\begin{equation}\label{div}
{\rm div}{\bf B}=0\,,
\end{equation}
\begin{equation}\label{e3}
{\rm curl}[{\bf v\,B}]=0\,.
\end{equation}

Here, the coordinates, density, velocity, and magnetic
field are expressed in units of their characteristic values
$R_0$, $\rho_0$, $v_0$, $B_0$, respectively; in addition,
\begin{equation}\label{e4}
\rho_0v_0^2=\frac{B_0^2}{4\pi}\,.
\end{equation}
The parameter
\begin{equation}\label{e5}
g=GM/\left(R_0 v_0^2 \right)\,,
\end{equation}
where $G$ the gravitational constant and $M$ the mass
of the central object.

Let us consider azimuthally symmetrical solutions
for these equations. The dimensioned variables will
be chosen as follows. Let us arbitrarily select a field
line, and adopt $R_0$ as the distance from the center
to the point where this line crosses the disk. We will
take $v_0$ to be the Keplerian velocity of a particle at
a distance $R_0$ from the center. We denote $2\,\pi\,\Psi_0$ to
be the magnetic flux through a circle with radius $R_0$
located in the disk and with its center on the axis
of symmetry. $B_0$ can then be defined by the relation
$ B_0=\Psi_0/ R_0^2$. The dimensioned density $ B_0=\Psi_0/ R_0^2$ is defined
by relation (8); $g=1$. Below, we will use only dimensionless
variables.

In the azimuthally symmetrical case, the field lines
form a two-parameter family of curves, which can be
presented in the form
\begin{equation}\label{e6}
{\bf R}(\alpha,\psi,\f)=\left(r(\alpha,\psi)\cos\phi,\,
r(\alpha,\psi)\sin\phi,\,z(\alpha,\psi)\right)\,,
\end{equation}
where $\phi=\f+\eta(\alpha,\psi)$. The parameters $\psi,\,\f$ specify
a field line, and $\alpha$ varies along a given line (${\bf B}\sim\p{\bf R}/{\p\alpha}$).
The substitution $\f\to\f-\f_0$ is equivalent
to a rotation by $\f_0$ of the field line as a whole.
The functions of two variables  $r(\alpha,\psi)$, $z(\alpha,\psi)$, and
$\eta(\alpha,\psi)$ are to be determined from the solution of the
MHD equations.
Following [31, 32], we will call the parameters $\alpha$, $\psi$,
and $\f$ the "frozen-in" coordinates. The cylindrical
coordinates $(r,\,\phi,\,z)$ can be expressed in terms of the
frozen-in coordinates by the formulas
\[
r=r(\alpha,\psi),\quad \phi=\f+\eta(\alpha,\psi),\quad z=z(\alpha,\psi).
\]
Let us make a transformation from Cartesian to
frozen-in coordinates in (4)–(7). $J$ denotes the Jacobian
of the transformation:
\begin{equation}\label{eq7}
J=\frac{\p(X,Y,Z)}{\p(\alpha,\psi,\f)} ={\bf R}_\alpha[{\bf R_\psi\,
 R_\f}]\,,
\end{equation}
where the subscripts denote differentiation with respect
to the corresponding variables
(${\bf R}_\alpha\equiv\p{\bf R}/\p\alpha,\,\dots$).
Since field lines do not cross, there
is a one-to-one relation between the frozen-in and
Cartesian coordinates, so that the Jacobian J never
vanishes. Note the following useful equalities, which
can be used to easily change between Cartesian and
frozen-in coordinates:
\begin{equation}\label{e_q1}
{\bf R}_\alpha\nabla=\frac\p{\p\alpha}\,,\quad
{\bf R}_\psi\nabla=\frac\p{\p\psi}\,,\quad
{\bf R}_\f\nabla=\frac\p{\p\f}\,.
\end{equation}
and also
\begin{equation}\label{e_q2}
\frac{{\bf R}_\alpha}{J}=[\nabla\psi,\nabla\f],\quad
\frac{{\bf R}_\psi}{J}=[\nabla\f,\nabla\alpha],\quad
\frac{{\bf R}_\f}{J}=[\nabla\alpha,\nabla\psi]\,,
\end{equation}
and
\begin{equation}\label{e_q3}
\frac{[{\bf R}_\psi,{\bf R}_\f]}{J}=\nabla\alpha,\quad
\frac{[{\bf R}_\f,{\bf R}_\alpha]}{J}=\nabla\psi,\quad
\frac{[{\bf R}_\alpha,{\bf R}_\psi]}{J}=\nabla\f\,.
\end{equation}
The operator $\nabla$ in frozen-in coordinates has the form
\begin{equation}\label{eq8}
\nabla=\frac1J
\left\{[{\bf R_\psi R_\f}]\frac\p{\p\alpha}
+[{\bf R_\f R_\alpha}]\frac\p{\p\psi}
+[{\bf R_\alpha R_\psi}]\frac\p{\p\f}\right\}\,.
\end{equation}
Suppose that the magnetic field strength is specified
by the equality
\begin{equation}\label{B}
{\bf B}=\frac{{\bf R}_\alpha}{J}=[\nabla\psi,\nabla\f]\,.
\end{equation}
In this case, (6) becomes an identity (it essentially
expresses the field in terms of the Euler potentials).
Expression (16) implies that we take the magnetic
flux (in units of $2\,\pi\,\Psi_0$) through a circle with radius r as
the frozen-in coordinate $\psi$. Note that the ratio ${\bf R}_\alpha/J$
is invariant with respect to the reparameterization
$\alpha\to\alpha'=g(\alpha,\psi,\f)$; i.e.,
\[
\frac{{\bf R}_\alpha}{J}=\frac{{\bf R}_{\alpha'}}{J'}\,,
\]
where $g(\alpha,\psi,\f)$ is an arbitrary function and
$J'={\p(X,Y,Z)}/{\p(\alpha',\psi,\f)} =J/g_\alpha$.
This can be used to
assign an arbitrary value to the Jacobian $J$ without
varyingthe field strength. It is convenient to relate $J$
with the plasma density:
\begin{equation}\label{rho}
J=1/\rho\,,
\end{equation}
in which case the magnetic field is
\begin{equation}\label{B1}
{\bf B}=\rho\,{\bf R}_\alpha.
\end{equation}

Let us introduce the three mutually orthogonal
unit vectors
\begin{equation}\label{es4}
{\bf e}^{(r)}=(\cos\phi,\sin\phi,0),\;\;
{\bf e}^{(\phi)}=(-\sin\phi,\cos\phi,0),\;\;
{\bf e}^{(z)}=(0,0,1)\,.
\end{equation}
We obtain from (10)
\begin{equation}\label{es5}
\left.
     \begin{array}{lcl}\ds
        {\bf R}_\alpha&=&r_\alpha\,{\bf e}^{(r)}+r\,\eta_\alpha\,{\bf e}^{(\phi)}+
        z_\alpha\,{\bf e}^{(z)}\,,\\
        {\bf R}_\psi&=&r_\psi\,{\bf e}^{(r)}+r\,\eta_\psi\,{\bf e}^{(\phi)}+
        z_\psi\,{\bf e}^{(z)}\,,\\
        {\bf R}_\f&=&r\,{\bf e}^{(\phi)}\,.
     \end{array}
\right\}
\end{equation}
Therefore, the Jacobian of the transformation is
\begin{equation}\label{J}
J=r\,(z_\alpha r_\psi-z_\psi r_\alpha)\,,
\end{equation}
and, in accordance with (17), the plasma density is
\begin{equation}\label{rho1}
\rho=\frac1{r\,(z_\alpha r_\psi-z_\psi r_\alpha)}\,.
\end{equation}

Let us expand the plasma velocity into three noncoplanar
vectors:
\[
{\bf v}=f{\bf R}_\alpha+\Omega{\bf R}_\f+\e{\bf R}_\psi=
\]
\begin{equation}\label{v}
=(fr_\alpha+\e r_\psi)\,{\bf e}^{(r)}+(fz_\alpha+\e z_\psi)\,{\bf e}^{(z)}+
r\,(f\eta_\alpha+\e\eta_\psi+\Omega)\,{\bf e}^{(\phi)}\,,
\end{equation}
where the functions $f,\,\Omega,\,\e$ may depend on the frozen-in
coordinates.
It turns out that, in both the stationary and nonstationary
cases [31, 32], Eqs. (5) and (7) in frozen-in
coordinates can be integrated in general form. Substituting
(23) and (17) into the continuity equation (5)
and transforming to frozen-in coordinates, we obtain
\begin{equation}\label{e2a}
\frac{\p f}{\p\alpha}+\frac{\p\Omega}{\p\f}+
\frac{\p\e}{\p\psi}=0.
\end{equation}
The electrical-field strength is
\begin{equation}\label{E}
{\bf E}=-\frac1c\,[{\bf vB}]=\frac1c\,(\e\,\nabla\f-\Omega\,\nabla\psi)\,.
\end{equation}
It follows that (7) in frozen-in coordinates has the
form
\begin{equation}\label{vr1}
{\rm curl}[{\bf v\,B}]=[\nabla\Omega\,\nabla\psi]-
[\nabla\e\,\nabla\f]=\rho\left(\frac{\p\Omega}{\p\alpha}\,{\bf R}_\f
+\frac{\p\e}{\p\alpha}\,{\bf R}_\psi
-\left(\frac{\p\Omega}{\p\f}+\frac{\p\e}{\p\psi}
\right){\bf R}_\alpha\right)=0\,.
\end{equation}
The derivative $\p\Omega/\p\f$ due to the assumed azimuthal
symmetry. Solving(24) and (26), we find that
$\e={\rm const}$, while the functions $f$ and $\Omega$ may depend
only on the single variable $\psi$. In this case, (5) and (7)
are identically satisfied.
The gradient$\nabla\psi$ has only $r$ and $z$ components; the
first term in (25) specifies the azimuthal component of
the electrical field:
\begin{equation}\label{Ephi}
 E_\phi=\frac{\e}{cr}\,.
\end{equation}
The velocity of the accretion disk ${\bf u}$ is also expressed
in frozen-in coordinates. Let us suppose that
the disk surface corresponds to the parameter $\alpha=0$;
i.e., $z(0,\psi)=0$. Then, the derivative $z_\psi(0,\psi)=0$.
The plasma density above the disk surface is
\begin{equation}\label{rho0}
\rho_0=\frac1{r\,z_\alpha r_\psi}\Big|_0\,,
\end{equation}
The $z$ component of the magnetic field is
\begin{equation}\label{Bz0}
B_{z0}=\rho\,z_\alpha\big|_0=\frac1{r\,r_\psi}\Big|_0\,.
\end{equation}
The radial component of the electrical field at the disk
surface is
\begin{equation}\label{Er0}
E_{r0}=-\frac1c\,[{\bf v\, B}]_r\big|_0=-\frac1{r_\psi c}\,
\left(\Omega+\e\eta_\psi\right)\big|_0\,,
\end{equation}
while the azimuthal component is defined by (27).
Using the frozen-in condition
\[
{\bf E}=-\frac1c\,[{\bf u\,B}]\,,
\]
the continuity of the $z$ component of the magnetic
field at the boundary between the plasma and disk,
and the continuity of the $r$ and $\phi$ components of the
electrical field, we obtain the components of the disk
velocity
\begin{equation}\label{ur_phi}
u_r=\e\,r_\psi\big|_0\,,\quad u_\phi=r\,(\Omega+\e\,\eta_\psi)\big|_0.
\end{equation}
After transforming to frozen-in coordinates, the
Euler equation (4) assumes the form
\[
\left(f\,\frac{\p}{\p\alpha}+\Omega\,\frac{\p}{\p\f}+\e\,\frac{\p}{\p\psi}
\right)(f{\bf R}_\alpha+\Omega{\bf R}_\f+\e{\bf R}_\psi)-
\frac{\p}{\p\alpha}\left(\rho\,{\bf R}_\alpha\right)=
\]
\begin{equation}\label{eu}
=-\frac12\,
\left\{[{\bf R_\psi R_\f}]\frac\p{\p\alpha}
+[{\bf R_\f R_\alpha}]\frac\p{\p\psi}
+[{\bf R_\alpha R_\psi}]\frac\p{\p\f}\right\}
\left(\rho\,{\bf R}_\alpha \right)^2-g\,\frac{\bf R}{R^3}\,.
\end{equation}
Substituting $\bf R$ the form (10) and $\rho$ in the form (22),
we can obtain equations for the functions $r(\alpha,\psi)$,
$z(\alpha,\psi)$, and $\eta(\alpha,\psi)$.

\section{BOUNDARY CONDITIONS}

Let us study the relation between the field components
and the parameters of the accretion disk,
assuming the disk is infinitely thin (which is valid for
a cool plasma). We will specify the volume density of
the disk in the form
\begin{equation}\label{eq1}
\rho_d(r,z)=\sigma(r)\,\delta(z)\,,
\end{equation}
where $\delta(z)$ is a delta function and $\sigma(r)$ is the mass
surface density. Equation (4) is valid for $z>0$ (or $z<0$);
i.e., beyond the disk. The dynamics of the plasma
in the disk are specified by the equation
\begin{equation}\label{e1d}
\rho_d({\bf u}\nabla){\bf u}+\rho({\bf v}\nabla){\bf v}=
-\frac12\,\nabla\,{\bf B}^2+({\bf B}\nabla){\bf B}
-(\rho_d+\rho)\,\frac{g\bf R}{R^3}\,.
\end{equation}

We will integrate (34) over $z$ in the small interval
$(-z_0,\,z_0)$ and take the limit $z_0\to+0$. This operation
will be applied to each term of the equation. In cylindrical
coordinates,
\begin{equation}\label{eq4}
\nabla={\bf e}^{(r)}\frac{\p}{\p r}+{\bf e}^{(z)}\frac{\p}{\p z}+
{\bf e}^{(\f)}\frac{\p}{r\p \phi}\,,
\end{equation}
so that
\[
\rho_d\,({\bf u}\nabla){\bf u}=\sigma(r)\,\delta(z)\,
\left(u_r\,\frac{\p}{\p r}+u_z\,\frac{\p}{\p z}+u_{\phi}\,\frac{\p}{r\p \phi}
\right)
({\bf e}^{(r)} u_r+{\bf e}^{(z)} u_z+{\bf e}^{(\phi)}u_{\phi})=
\]
\begin{equation}\label{eq5}
=\sigma(r)\,\delta(z)\left[\left(u_r\frac{\p u_r}{\p r}-
\frac{u_{\phi}^2}{r} \right){\bf e}^{(r)}
+u_r\left(\frac{\p u_{\phi}}{\p r}+\frac{u_{\phi}}{r} \right)
{\bf e}^{(\phi)}\right].
\end{equation}
When deriving(36), we took into account azimuthal
symmetry, the equality $u_z=0$, and the relations
$\p{\bf e}^{(r)}/\p\phi ={\bf e}^{(\phi)}$,
$\p{\bf e}^{(\phi)}/\p\phi=-{\bf e}^{(r)}$.
Integrating (36), we obtain
\begin{equation}\label{eq6}
\int\limits_{-z_0}^{z_0}\!\rho_d\,({\bf u}\nabla){\bf u}\,dz
=\sigma(r)\left[\left(u_r\frac{\p u_r}{\p r}-
\frac{u_{\phi}^2}{r} \right){\bf e}^{(r)}
+u_r\left(\frac{\p u_{\phi}}{\p r}+\frac{u_{\phi}}{r}
 \right){\bf e}^{(\phi)}\right].
\end{equation}

If the magnetic field and the plasma velocity above
the accretion disk (z > 0) are known, then the solution
in the region under the disk continues in accordance
with the rules
\begin{equation}\label{eq1b}
B_z\to B_z\,,\quad B_r\to -B_r\,,\quad B_{\phi}\to-B_{\phi}\,.
\end{equation}
\begin{equation}\label{eq1v}
v_z\to-v_z\,,\quad v_r\to v_r\,,\quad v_{\phi}\to v_{\phi}\,.
\end{equation}
Since the field components $B_{r}$, $B_{\phi}$ at the equator
(for $z=+0$), this implies that the field line is discontinuous
there.

Using relations (39), it can be shown that, in the
\[
\int\limits_{-z_0}^{z_0}\!\rho\,({\bf v}\nabla){\bf v}\,dz=0\,.
\]
Next, we obtain
\begin{equation}\label{eq7a}
\int\limits_{-z_0}^{z_0}\!\nabla B^2\,dz=0\,,
\end{equation}
since $B^2$ is a continuous function of $z$.

The term $({\bf B}\nabla){\bf B}$ will be written in the form
\begin{equation}\label{eq8a}
({\bf B}\nabla){\bf B}=
\left(B_r\frac{\p}{\p r}+B_z\frac{\p}{\p z}+B_{\phi}\frac{\p}{r\p \phi}
\right)
\left(
B_r{\bf e}^{(r)}+B_z{\bf e}^{(z)}+B_{\phi}{\bf e}^{(\phi)}
\right).
\end{equation}
If the integration is taken over an infinitely small
interval, a nonzero contribution will be obtained only
from terms containingth e derivative $\p/\p z$ (a derivative
of a discontinuous function yields a $\delta$ function).
After the integration, we obtain
\begin{equation}\label{eq9}
\int\limits_{-z_0}^{z_0}\!({\bf B}\nabla){\bf B}\,dz=
2\,B_z\left(B_r{\bf e}_r+B_{\phi}{\bf e}_{\phi}\right).
\end{equation}
The integration of the last term in (34) is trivial due to
the presence of the $\delta$ function:
\begin{equation}\label{eq10}
\int\limits_{-z_0}^{z_0}\!\frac{\rho_d\, g\,{\bf R}}{R^3}\,dz=
\frac{\sigma\,g}{r^2}\,{\bf e}^{(r)}.
\end{equation}
Collecting these various relations, we obtain two
equalities (for the $r$ and $\phi$ components in the equation),
which are a direct consequence of (34) in the
model with an infinitely thin disk:
(\ref{e1d}) â ìîäåëè áåñêîíå÷íî òîíêîãî äèñêà:
\begin{equation}\label{eq11}
u_r\,\frac{\p u_r}{\p r}-\frac{u_{\phi}^2}{r}=
\frac2{\sigma}\,B_rB_z-\frac{g}{r^2}\,,
\end{equation}
\begin{equation}\label{eq12}
\frac{u_r}r\,\frac\p{\p r}(r\, u_{\phi})=
\frac2{\sigma}\,B_{\phi}B_z\,.
\end{equation}
The conservation of mass yields another relation:
\begin{equation}\label{eq13}
\frac1r\,\frac\p{\p r}(r\,\sigma u_r)+2\,j_z=0\,,
\end{equation}
where $j_z=\rho\,v_z$ is the density of the $z$ component
of the mass flux. In (44)–(46), the magnetic-field
components and plasma parameters are taken for the
disk surface ($z=+0$).
Equations (44)-–(46) play the role of boundary
conditions, which must be satisfied by the solution.
They were derived from the equations of motion in
differential form. Obviously, the same conditions can
be obtained using the integral form of the equations of
motion.

\section{SELF-SIMILAR SOLUTIONS}

We are primarily concerned here with the plasma
dynamics immediately above the disk, at distances
$z$ much smaller than the radius of the disk $R_{\rm disk}$.
In the limit $z \ll R_{\rm disk}$, only two parameters with the
dimensions of length remain in the problem: $z$ and $r$.
Therefore, the solution will be self-similar in this limit
[11]. Let us underscore an important feature of these
solutions: they describe flows only at small distances
from the disk, and are not applicable at distances
comparable to or exceeding the size of the disk.
Equations (4)–-(7) have self-similar solutions of
the form .
\begin{equation}\label{au}\left.
\begin{array}{rcl}\ds
{\bf v}(r,z,\phi)&=&r^{-\delta_v}\t{\bf v}(z/r,\phi)\,,\\
\rho(r,z)&=&r^{-\delta_{\rho}}\t\rho(z/r)\,,\\
{\bf B}(r,z,\phi)&=&r^{-\delta_B}\t{\bf B}(z/r,\phi)\,.
\end{array}
\right\}
\end{equation}

The superscripts $\delta_v$, $\delta_{\rho}$, and $\delta_B$ are determined from
the following conditions. Substituting (47) into (4)
leads to the equations
\begin{equation}\label{ind1}
2\delta_B-\delta_\rho=2\delta_v=1\,.
\end{equation}
Equality (27) is satisfied under the condition
\begin{equation}\label{ind2}
\delta_v+\delta_B=1\,.
\end{equation}
Thus,
\begin{equation}\label{delta}
\delta_v=\delta_B=\frac12\,,\quad \delta_\rho=0
\end{equation}
and hence,
\begin{equation}\label{auto}\left.
\begin{array}{rcl}\ds
{\bf v}(r,z,\phi)&=&r^{-1/2}\,\t{\bf v}(z/r,\phi)\,,\\
\rho(r,z)&\,=\,&\t\rho(z/r)\,,\\
{\bf B}(r,z,\phi)&=&r^{-1/2}\,\t{\bf B}(z/r,\phi)\,.
\end{array}
\right\}
\end{equation}
Note that it is possible to determine the subscripts
unambiguously only for $u_r\ne 0$; i.e., when there is
advection of the plasma toward the center of gravitation.
When $u_r=0$, one of the subscripts is arbitrary,
so that there is a single-parameter family of selfsimilar
solutions. A large number of such solutions
have been studied previously (see, for example, [28]
and references therein). For clarity, we will compare
our results with those of [16], which is similar to our
study in terms of the formulation for the wind. The
case $\delta_\rho=3/2$, $\delta_B=5/4$. was considered in [16].

In frozen-in coordinates, the law of similarity (51)
is described by the relation
\begin{equation}\label{au1}
{\bf R}(\alpha,\psi,\f)=\psi^{2/3}\,
\t{\bf R}\left(\frac\alpha\psi\,,\,\f\right)\,,
\end{equation}
Therefore, the two-variable functions in (10) are expressed
in terms of one-variable functions. Denoting
$s=\alpha/\psi$, we can write
\begin{equation}\label{au2}
r(\alpha,\psi)=\psi^{2/3}\,\t r(s)\,,\;\;
z(\alpha,\psi)=\psi^{2/3}\,\t z(s)\,,\;\;
\eta(\alpha,\psi)=\t\eta(s)\,.
\end{equation}

Let us verify the equivalence of (51) and (53). Substituting(53)
into (22), we derive the plasma density:
\begin{equation}\label{au3}
\rho(s)=\frac32\,\frac1{\t r(\t r\t z_s-\t z\t r_s)}\,,
\end{equation}
where the subscript $s$ denotes differentiation with
respect to $s$. This function depends only on $s$, or,
if we switch to cylindrical coordinates, on the ratio
$z/r$, as in (51). The components of the magnetic-field
strength are
\begin{equation}\label{au4}
B_r=\psi^{-1/3}\,\rho\,\t r_s\,,\;\;B_z=\psi^{-1/3}\,\rho\,\t z_s\,,\;\;
B_\phi=\psi^{-1/3}\,\rho\,\t r\t \eta_s\,.
\end{equation}
We find for the components of the plasma velocity
\begin{eqnarray}\label{au5}
v_r&=&\psi^{-1/3}\left(\frac23\,\e \t r+(f-\e s)\t r_s\right),\\
v_z&=&\psi^{-1/3}\left(\frac23\,\e \t z+(f-\e s)\t z_s\right),\\
v_\phi&=&\psi^{-1/3}\,\t r\left(\psi\Omega+(f-\e s)\t \eta_s\right).
\end{eqnarray}
The law of similarity ${\bf v}\sim r^{-1/2}$ will be satisfied if
$f=\rm const$, while $\Omega$ depends on $\psi$ as follows:
\begin{equation}\label{au6}
\Omega(\psi)=\t\Omega/\psi\,.
\end{equation}
represents the angular velocity of the selected field
line in units of the Keplerian angular velocity.
Substituting(52) into (32), we obtain a system
of three ordinary second-order differential equations\footnote{%
The equations are cumbersome, and we do not present them
here. The analytical manipulations were performed using the
Maple package.}
for the functions $\t r(s)$, $\t z(s)$, and $\t \eta(s)$. Since second
derivatives appear in the equations in linear form, the
system can be solved for the second derivatives. For
$\e\ne0$, the right-hand sides of the equations depend
explicitly on $s$; $f$ and $s$ appear in the equations only in
the combination $f-\e s$. The parameter $s$ has an arbitrary
zero point. Indeed, shifting the reference point
for $s$ via the substitution $s=s'+s_0$ is equivalent to
redefining the constant $f$: $f\to f'=f-\e s_0$. We will
take the surface of the accretion disk as the reference
point for $s$: $\t z(s=0)=0$.

The resulting equations can be written as a system
of six ordinary first-order differential equations.
We introduce the six-dimensional vector $\xi_i$ with the
components
\begin{eqnarray}\label{au7}
\xi_1=\t r_s\,,\;\;&\xi_2=\t z_s\,,\;\;&\xi_3=\t\eta_s\,,\\
\xi_4=\t r\,,\;\;&\xi_5=\t z\,,\;\;&\xi_6=\t \eta\,.
\end{eqnarray}
Then, for $i=1,\,2,\,3$ the equations have the structure
\begin{equation}\label{au8}
\frac{d\xi_{1,2,3}}{ds}=\frac{N_{1,2,3}}{D}\,,
\end{equation}
while for $i=4,\,5,\,6$
\begin{equation}\label{au9}
\frac{d\xi_{4,5,6}}{ds}=\xi_{1,2,3}\,.
\end{equation}
The denominator $D$ is the determinant of a certain
$3\times3$ matrix; it appears when the initial system of
equations is solved relative to the highest derivatives.
The functions $N_i$ and $D$ depend on $\xi_1,\dots,\,\xi_5$; $\xi_6$ does
not appear in the equations due to the azimuthal
symmetry of the problem. In addition, the right-hand
sides of (62) depend on $s$
.
For fixed $\psi$ and $\f$, the curve ${\bf R}=\psi^{2/3}\,\t{\bf R}(s,\f)$
specifies a magnetic-field line in parametric form. In

the case of self-similar solutions, it will suffice to find
one field line; others can be derived using the similarity
transformation (multiplication by $\psi^{2/3}$) and
rotation about the $z$ axis by some angle $\f$. The field
line corresponding to the frozen-in coordinate $\psi=1$
crosses the disk at the distance $r(0)=1$. The value
$\t\eta(0)$ can be arbitrary, for example, zero. The solution
of the system of differential equations should then
satisfy the initial conditions
\begin{equation}\label{ini1}
\t r(0)=1\,,\quad \t z(0)=0\,,\quad \t\eta(0)=0\,.
\end{equation}
The frozen-in coordinate $\psi$ and cylindrical coordinate
$r$ are related by the simple expression
\begin{equation}\label{rpsi}
r=\psi^{2/3}\,.
\end{equation}
To obtain an unambiguous solution of the system of
equations, we must have another three conditions.

It follows from (54) that the plasma density at the
disk surface is $\rho=3/(2\,\t z_s^{(0)})$
Therefore, the components of the magnetic field in the disk are
\begin{equation}\label{c4}
B_r^{(0)}=\frac32\,\frac{\t r_s^{(0)}}{\t z_s^{(0)}}\,\psi^{-1/3}\,,
\quad B_z^{(0)}=\frac32\,\psi^{-1/3}\,,\quad
B_\phi^{(0)}=\frac32\,\frac{\t \eta_s^{(0)}}{\t z_s^{(0)}}\,\psi^{-1/3}\,,
\end{equation}
where the superscript 0 denotes values for $z=+0$.

Let us write the boundary conditions (44)–(46)
for the self-similar solution. Substituting(65) into
(31) and applying the initial conditions, we obtain the
velocity of the accretion disk:
\begin{equation}\label{uaccr}
u_r=\frac23\,\e\,\psi^{-1/3}\,,\quad u_\phi=\t\Omega\,\psi^{-1/3}\,.
\end{equation}
The velocity component $u_r<0$, so that $\e$ should also
be negative. Together with the plasma velocity, $u_r$ and
$u_\phi$ depend on $r$ as $r^{-1/2}$.
It follows from (44) or (45) that the mass surface
density of the disk is a linear function of $r$:
\begin{equation}\label{mu}
\sigma=\mu\,r \,,
\end{equation}
where $\mu={\rm const}$. In (46), $j_z$ is expressed using (54)
and (56) as follows:
\begin{equation}\label{c6}
j_z=\frac32\,f\,\psi^{-1/3}\,.
\end{equation}
We derive from the continuity equation (46)
\begin{equation}\label{c7}
\mu=3\,f/|\e|\,.
\end{equation}

Substituting(66), (67), and (68) into the boundary
conditions (44) and (45), we obtain
\begin{equation}\label{c8}
\t\Omega^2=1-\frac29\,\e^2+\frac{3\,\e}{2\,f}\,\frac{\t r_s^{(0)}}
{\t z_s^{(0)}}\,,
\end{equation}
\begin{equation}\label{c9}
\t \eta_s^{(0)}=-\frac29\,f\t\Omega\,\t z_s^{(0)}\,.
\end{equation}
This last expression can be used to calculate the
azimuthal component of the magnetic field at the
equator:
\begin{equation}\label{c8a}
B_\phi^{(0)}=-\frac13\,f\,\t\Omega\,\psi^{-1/3}\,.
\end{equation}
We can see from these conditions that the rotational
velocity of a field line is lower than the Keplerian
velocity. Only in the limit $\e\to0$ do we obtain $\t\Omega=1$,
which corresponds to a Keplerian accretion disk. Note
that the boundary conditions in the disk fully specify
the azimuthal magnetic field above the disk. Since
$\e$ does not appear in (72) and (73), these equalities
should be satisfied even in the limit $\e\to0$.

The conditions (44) and (46) make it possible to
relate in a self-consistent way all the parameters of
the accretion disk and of the outflowing plasma, and,
in particular, to take into account the decrease in the
surface density of the disk due to the expansion of the
plasma. The equality (45) leads to the additional condition
(72), which must be satisfied by the solution.

\section{SOLUTION IN THE VICINITY OF THE DISK}

Let us consider an important limiting case, when
the radial velocity of the disk is small compared to the
Keplerian velocity, $|\e|\ll 1$. Let us also assume that
the velocity of the plasma outflow from the disk is
fairly low. Then, the following equations are valid in
the vicinity of the disk:
\begin{equation}\label{eqv1}
\frac{d\t r_s}{ds}=\t r_s\,G\,,\quad
\frac{d\t z_s}{ds}=\t z_s\,G\,,\quad
\frac{d\t \eta_s}{ds}=\t \eta_s\,G\,.
\end{equation}
Here,
\begin{equation}\label{eqv2}
G=\frac1{6\,f^3}\,\frac{6\,f\,\t z_s(3\t r_s\,\delta\t r-
\t z_s\,\t z)+9\,\e\,\t r_s^2-2\,\e\, f\,\t z_s\,\t \eta_s}
{\t z_s({\t r_s}^2+{\t z_s}^2+{\t \eta_s}^2)}\,,
\end{equation}
where $\delta\t r=\t r-1$. To derive (74), $\t\Omega$ in the form (71)
is substituted into the exact equations (62) and expanded
in small parameters. Equations (74) are valid
when $\delta\t r$ and $\t z$ are small compared to unity and
\begin{equation}\label{cond}
f^2\t r_s\ll1\,,\quad f^2\t z_s\ll1\,,\quad f^2\t \eta_s\ll1\,.
\end{equation}
These conditions follow from the requirement that
subsequent terms of the expansion be small. It follows
from (74) that
\begin{equation}\label{eqv3}
\xi_s=\xi_s^{(0)}\exp\Big(\int\limits_0^s\!G\,ds\Big)\,,
\end{equation}
where $\xi_s$ is any one of the three functions $\t r_s$, $\t z_s$,  or
$\t \eta_s$. After integrating (77) with the initial conditions
$\t z(0)=0$, $\delta\t r(0)=0$, $\t \eta(0)=0$, we find that a field line
is straight in the vicinity of the disk:
\begin{equation}\label{eqv4}
\delta\t r(s)=k_r\,\t z(s)\,,\quad \t \eta(s)=k_\eta\,\t z(s)\,,
\end{equation}
where $k_r$ and $k_\eta$ are constant. As we can see from
(72), $k_\eta=-2\,f/9$ in the approximation considered.
$k_r$ specifies the direction of the poloidal field in the
vicinity of the disk.
Using equality (78), the second of Eqs. (74) can be
presented in the form
\begin{equation}\label{eqv5}
\frac{d^2\t z}{ds^2}=w^2\,\t z-|\e|\,\zeta\,,
\end{equation}
where
\begin{equation}\label{eqv6}
w^2=\frac1{f^2}\,\frac{3\,k_r^2-1}{k_r^2+k_\eta^2+1}\,,
\end{equation}
\begin{equation}\label{eqv7}
\zeta=\frac{1}{6\,f^3}\,\frac{9\,k_r^2-2\,f\,k_\eta}{k_r^2+k_\eta^2+1}\,.
\end{equation}
When $w^2<0$, (79) has an oscillating solution, which
is physically unacceptable. The inequality $k_r^2>1/3$,
obtained in [16], must be satisfied by field lines moving
away from the disk. This implies that a magnetic-field
line must be inclined to the disk at an angle smaller
than $60^\circ$.

Unlike [16], we also obtained another constraint.
When $w^2>0$, the solution (79) has the form
\begin{equation}\label{eqv8}
\t z=\frac{|\e|\,\zeta}{w^2}\left(1-\cosh(ws)\right)+
\frac{\t z_s^{(0)}}{w}\,\,\sinh(ws)\,,
\end{equation}
where $\t z_s^{(0)}$  is an arbitrary constant equal to the derivative
$\t z_s$ in the disk. It follows from (82) that solutions
cannot exist for arbitrarily small $\t z_s^{(0)}$. The field line will
move away from the disk only when
\begin{equation}\label{eqv9}
\t z_s^{(0)}>\t z_{s\,\min}=\frac{|\e|\,\zeta}{w}\,.
\end{equation}
This is equivalent to a constraint on the $z$ component
of the plasma outflow velocity:
\begin{equation}\label{eqv9a}
v_z\big|_{z=0}>v_{z\,\min}=f\,\frac{|\e|\,\zeta}{w}\,.
\end{equation}
The nature of this constraint, which appears when
$\e\ne 0$, is easy to understand. In this case, each particle
of the plasma moves along its orbit with a velocity
that is smaller than the Keplerian velocity. In the $z$
direction, the particle is in the effective potential
\[
U(\t z)=-\frac12\,w^2\,\t z^2+|\e|\,\zeta\,\t z\,.
\]
In order to be ejected from the disk, the particle must
overcome the potential barrier corresponding to the
difference between the gravitational and centrifugal
forces.

The plasma density in the vicinity of the disk is
\begin{equation}\label{eqv10}
\rho=\frac{3\,}{2\,\t z_s}=\frac3{2\left(z_s^{(0)}{\rm ch}(ws)-
\ds\frac{|\e|\,\zeta}{w}\,{\rm sh}(ws)\right)}\,.
\end{equation}
In the disk itself, it is finite:
\begin{equation}\label{eqv11}
\rho\big|_{z=0}=\frac{3\,}{2\,\t z_s^{(0)}}
<\frac{3\,w}{2\,|\e|\,\zeta}\,.
\end{equation}
In the limit $\e\to 0$, $\t z_s^{(0)}\to 0$, the density in the disk
increases without bound, as in [16].

\section{PASSAGE THROUGH CRITICAL SURFACES}

The general solutions of the equations describing stationary
MHD flows are singular on certain
surfaces, usually called critical surfaces. These have
already been found for one-dimensional flows, as was
proposed by Parker for the solar wind [33]. In one dimensional
solutions, they are manifest as critical
points, while, in the three-dimensional case, a set of
critical points forms a surface. In the one-dimensional
solutions, the critical points coincide with sonic
points, where the local velocity of weak perturbations
becomes comparable to the flow velocity, while the
type of equation changes from elliptical to hyperbolic
and back. Even the earliest solutions obtained for
axisymmetrical flows indicated that this is not the
case in general [16]. The nature of critical surfaces
and their role in the formation of stationary MHD
solutions was investigated in [34, 35]. It turns out
that critical surfaces divide regions of solutions with
different cause–effect relations, and these surfaces do
not, in general, coincide with the surfaces where the
form of the equation changes.

Critical surfaces as applied to self-similar solutions
were studied in [36]. The parameters of critical
surfaces for self-similar flows with $E_{\varphi}\ne 0$ were
studied in detail in [27]. In these solutions, the locations
of the critical surfaces are specified by the
zeros of the denominator D in (62). In the general
case (with a nonzero wind temperature), the denominator
vanishes at the slow magnetoacoustic separatrice
surface, Alfven surface, and fast magnetoacoustic
surface. Regularizing the solution on these
surfaces makes it possible to determine the density of
the plasma flux from the disk and the two tangential
components of the magnetic field at the disk surface
(the normal component is specified by the conditions
adopted for the problem). In the case of a cool plasma,
the slow magnetoacoustic velocity is zero, and one
of the regularization conditions disappears. Therefore,
the plasma flux from the disk surfacemust be specified
for a cool plasma. This means that the ratio f of
the density of the plasma flux from the disk and the
z component of the magnetic field is also specified.
A similar approach was used in [16] (where $k$ is analogous
to $f$).

Thus, in self-similar flows of cool plasma, regularizing the
solution on the Alfven and fast magnetoacoustic
critical surfaces specifies both the azimuthal
and tangential components of the magnetic
field based on the flow. One characteristic feature of
self-similar solutions is that, as a rule, it is not possible
to regularize them on the fast magnetoacoustic
critical surface, only at the Alfven surface. Recently,
Vlahakis et al. [37] were able to regularize a selfsimilar
solution on the fast magnetoacoustic surface,
but, even in this case, the solution disappears very
rapidly behind this surface. Therefore, as in [16], we
will not try to regularize the solution on the fast magnetoacoustic
surface. As a result, one of the components
of the magnetic field (the azimuthal component
in [16]) remains free and must be specified a priori.
The other component is determined from the solution
and the regularization condition at the Alfven surface.
The denominator $D$ in (62) contains the factor
\begin{equation}\label{au10}
e_1\equiv\rho-(f-\e s)^2\,.
\end{equation}
In the disk, $e_1>0$ when $f^2\t z_s^{(0)}<3/2$. Since the density
decreases along a field line, $e_1$ vanishes at some
point on the Alfven surface. To keep the solution finite,
the functions Ni must also vanish at this point. After
substituting $\rho=(f-\e s)^2$ into $N_i$, all the functions
$N_i$ possess the common factor
\[
e_2\equiv \left(-108\,\t\Omega\,b+ \left(12\,{\t
r}^{2}\e\,\t\Omega\,\t z-27\, \t\eta_s \right) b^2 \right) \t
r_s^2+
\]
\[
+ \left(\left( 8\,\t r^3\t z\e^2\t \eta_s+ 8\,\t r\t
z^3{{\e}}^{2}\t \eta_{{s}}+12 \,\t r{\t z}^{2}{\e}\,\t\Omega\,\t
z_{{s}}-12\,{\t r}^{3}{\e}\,\t\Omega\,\t z_{{s }}+36\,{\t r}^{3}\t
z{\t\Omega}^{2}\t \eta_{{s}} \right) {b}^{2}-108\,{\frac {
\t\Omega\,b\t z\t z_{{s}}}{\t r}} \right) \t r_{{s}}+
\]
\[
+ \left( -36\,{\t r}^{4}\t z_{{s}}{
\t \Omega}^{2}\t \eta_{{s}}-8\,\t {r}^{4}\t z_{{s}}{{\t \e}}^{2}\t \eta_{{s}}-12\,{\t r
}^{2}\t z{\e}\,\t \Omega\,{\t z_{{s}}}^{2}-8\,\t {r}^{2}\t z_{{s}}{{\e}}^{2}
\t \eta_{{s}}{\t z}^{2}-27\,{\t r}^{2}{\t \eta_{{s}}}^{3}-27\,\t \eta_{{s}}{\t z_{{s}}}^
{2} \right) {b}^{2}+
\]
\begin{equation}\label{au11}
+54\,{\frac {\t \eta_{{s}}}{\sqrt {{\t r}^{2}+{\t z}^{2}}}}-
108\,b\t \Omega\,{\t r}^{2}{\t \eta_{{s}}}^{2}\,,
\end{equation}
where $b=f-\e s$. Therefore, the condition that the
denominator and three functions $N_i$ in (62) simultaneously
vanish results in the system of two equations
\begin{equation}\label{au12}
e_1=0\,,\quad e_2=0\,,
\end{equation}
which can be solved analytically for $\t r_s$ and $\t z_s$. Consequently,
the derivatives $(\t r_s,\,\t z_s)$ can be taken to be
specified at the critical point $(\t r,\,\t z)$.

\section{SOLUTION IN THE LIMIT $\e \to 0$.}

It is of interest to consider the solution in the
limit of very slow accretion, when $\e \to 0$. This case
is of interest because the disk becomes Keplerian in
this limit ($\t\Omega =1$), so that our results can easily be
compared to those of [16], and the physics of accretion
made possible by a wind outflow becomes particularly
simple.

In the case of small $\e$, $\e s$ can be neglected compared
to $f$. The parameter $s$ then disappears from the
right-hand sides of (62), and the simplified equations
have two integrals of motion: the momentum $L$ and
the energy $W$. For the self-similar solution,
\begin{equation}\label{mom}
L=\t r^2 \left(f\,\t\Omega-\t\eta_s\,(\rho-f^2)\right)\,,
\end{equation}
\begin{equation}\label{ener}
W=\frac12 f^2\left(\t r_s^2+\t r^2\t \eta_s^2+\t z_s^2 \right)
-\frac12\,\t\Omega^2 \t r^2-\frac{1}{\sqrt{\mathstrut\t r^2+\t z^2}}\,,
\end{equation}
where we can assume $\t\Omega=1$ on the right-hand sides
of the equality.

Substituting the values at the disk surface into
(90) and using (64) and (72), we find
\begin{equation}\label{mom1}
L= \frac43\,f\left(1-\frac16\,f^2\t z_s^{(0)}\right)\,.
\end{equation}
Since $\t z_s^{(0)}\ge0$, $L\le L_{\max}=\frac43\,f$.
This new important
constraint on the angular-momentum flux carried
away from the disk follows from the boundary conditions
at the disk. It means that a relatively small flux of
angular momentum must be carried away by the wind
to enable accretion. In addition, the boundary conditions
at the disk specify the azimuthal component of
the magnetic field, and the problem is fully specified
even when the solution is regularized at one Alfven
point.

As we can see from (84), in the limit $\e \to 0$, the
initial velocity of the plasma outflow can be arbitrarily
small. In this case, the energy tends to the minimum
$W_{\min}=-3/2$ [see (91)]. Assuming that the conditions
(76) are satisfied at the disk surface, we obtain
$L=4/3\,f$, $W=-3/2$. Therefore, at the Alfven point,
\begin{equation}\label{alv1}
  \t r^2=\frac Lf=\frac43.
\end{equation}
The energy integral and Eqs. (89) can be used to
express analytically the derivatives $\t r_s$, $\t z_s$, and $\t\eta_s$ at
the Alfven point in terms of $\t z$. A straightforward but
fairly cumbersome analysis shows that the solutions
(89) exist for $L=4/3\,f$, $W=-3/2$ if
\begin{equation}\label{alv2}
f>\frac9{16}\,\sqrt{6(5+3\sqrt{3})}\approx4.4\,.
\end{equation}
$\t z$ must be confined within the interval
\begin{equation}\label{alv3}
z_{\,\min}\le \t z\le z_{\,\max}\,,
\end{equation}
where
\begin{equation}\label{alv4}
z_{\,\max}=\left[\frac{216f\sqrt{16f^2-45\mathstrut}-736f^2-1215}
{1200f^2}\right]^{1/2}\,,
\end{equation}
$z_{\,\min}=0$ when $f<8.27$, and
\begin{equation}\label{alv5}
z_{\,\min}=\frac2{\sqrt{3}}\,\frac{\sqrt{\mathstrut2351f^4-159894f^2-59049}}
{329f^2+243}
\end{equation}
when $f>8.27$. It follows from (96) that $\t z<0.327$
at the Alfven point; this means that the Alfven point
cannot be far from the disk surface.

We solved (62) and (63) in the limiting case $\e\to0$
numerically using standard techniques. We started
from the Alfven point, with $f$ as the input parameter.
As the initial condition, we took the Alfven point with
$\t r=2/\sqrt{3}$ and $\t z$ from the interval (95); the derivatives
were selected using the described procedure. At the
Alfven point, the right-hand sides of (62) were determined
using l'Hopitale's rule. The initial value for
$\t z$ was selected so that the field line crossed the disk
at the point $\t r=1$. We adopted a value slightly larger
than $-3/2$ for $W$. The reason for this is that, when
$W=-3/2$, the plasma velocity at the disk surface
vanishes and the plasma density at the disk displays
a nonintegrable singularity: $\rho=3/(2\,w\,\t z)$. Therefore,
we chose W so that .$\t z_s^{(0)}\sim 10^{-3}$ in the disk, or, in
other words, so that the velocity of the plasma outflow
relative to the disk was of the order of $f\cdot 10^{-3}$ of
the Keplerian velocity. The requirement that $\t z_s^{(0)}$  be
nonzero also results from the fact that the inequality
(84) should be rigorous.

Figure 4 presents the $f$ dependence of $\t z$ at the
Alfven point together with $z_{\,\min}$ and $z_{\,\max}$. The minimum
$f$ for which the solution exists is $f_{\min}=5.28$.
$z_{\,\max}=z_A$ when $f=f_{\min}$.

Figure 5 presents a line of the poloidal magnetic
field for $f=6$, which does not differ much from $f_{\min}$.
$k_r=0.96$, so that the line is inclined to the equator
at an angle that is close to $\pi/4$; at the disk surface,
$\t z_s=10^{-3}$. There is no significant acceleration of
the plasma; at the Alfven point, $\t z_s=0.033$ and $\t r_s=0.010$,
so that the components of the plasma velocity
$v_z$ and $v_r$ are small compared to the Keplerian velocity.
Figure 6 presents the plasma density on a field line
for various heights. In the case of motion from the disk
toward the Alfven point, $s$ varies from 0 to $s_{\,\max}\approx 30$.
Therefore, $\e s$ can be neglected compared to $f$ when
\begin{equation}\label{cond1}
  |\e|\ll f/s_{\,\max}\,.
\end{equation}
This condition, as well as the inequality (76), ensures
the applicability of the approximation used in the
case considered. The analytical solution (82) coincides
with the numerical solution along nearly the
entire field line, with the exception of the immediate
vicinity of the Alfven point.

\section{DISCUSSION}

The conclusion that the plasma outflow from an
accretion disk can exert a substantial effect on the
dynamics of the accretion disk is not new [19]. The
originality of our work is our consideration of the case
when plasma falls onto the gravitating center along
with the frozen-in magnetic field, rather than seeping
across magnetic-field lines (see, for example, [21] and
references therein). Our results indicate that dissipationless
disk accretion of the type considered is,
indeed, possible, as is demonstrated by the derived
self-consistent solution.
The mechanism for carrying away angular momentum
that we have considered should be taken into
account in studies of accretion in sources displaying
violent ejections of matter. This is particularly important
for our understanding of the processes occurring
in a number of peculiar sources, prominently represented
by SS433 [2]. Objects in which disk accretion
occurs primarily due to the carrying away of angular
momentum by an outflowing wind can have bolometric
luminosities that are appreciably lower than
the kinetic-energy flux in the outflowing wind. This
is what leads to the most important difficulties in
explaining the plasma-ejection mechanism in both
SS433 and young stellar objects.
The derived self-consistent solution possesses a
number of interesting properties. The boundary conditions
(71), (72) fully specify the azimuthal magnetic
field in the disk. In this case, with fixed $f$ and
$L$, the toroidal and tangential components of the
magnetic field are determined unambiguously, unlike
in the study of Blandford and Payne [16]. The flow
turns out to be fully specified. This begs the question
of whether there exists the fundamental possibility
of passing through the fast magnetoacoustic critical
point, since, at first glance, no free parameters remain.
In fact, there is still one free parameter: $\e$, which specifies
the radial velocity of the plasma in the disk. This
parameter can be used to regularize the solution at the
fast magnetoacoustic critical point. If this is possible,
the entire solution will prove to be unambiguously
specified.
An unexpected feature of the obtained solutions is
that, in the limiting case of dissipationless accretion,
100\% of the infalling matter is ejected back out from
the disk. In this respect, our results are similar to
those of Lery et al. [38], who also considered dissipationless
accretion, but without the formation of a
disk. Both of these results indicate that, in any case,
dissipation is probably unavoidable, to provide for the
infall of some fraction of the matter onto the gravitating
center. On the other hand, this feature of the
flow may well be due to its self-similarity. Additional
studies are required before we will be able to draw final
conclusions.
A natural extension of our study is the consideration
of accretion due simultaneously to matter outflow
from the disk and dissipation. Even at this stage,
it should be possible to determine the relationship
between the disk luminosity and the flux of outflowing
matter, which can be compared to observations.

\section{ACKNOWLEDGMENTS}

The authors thank G.S. Bisnovatyi-Kogan,
V.M. Chechetkin, and G. Pelletierre for useful discussions.
We also acknowledge Johnatan Ferreira,
whose numerous remarks on the connection between
disk accretion and jets initiated this study. This work
was supported by the Russian Foundation for Basic
Research (project no. 03-02-17098), joint INTASESA
grant 99-120 and the program of the Presidium
of RAS "Non-stationary phenomena in astronomy."

\vspace{5mm}
REFERENCES
\vspace{5mm}

1. C. M. Urry and P. Padovani, Publ. Astron. Soc. Pac.
107, 803 (1995).

2. A. M. Cherepashchuk, Itogi Nauki Tekh. 38, 60
(1988).

3. M. Livio, Phys. Rep. 311, 225 (1999).

4. I. F. Mirabel and L. F. Rodrigu es, Nature 392, 673
(1998).

5. M. C. Weisskopf, J. J. Hester, F. A. Tenant, et al.,
Astrophys. J. 536, L81 (2000).

6. S. V. Bogovalov and D. V. Khangoulian, Mon. Not. R.
Astron. Soc. 336, L53 (2002).

7. N. I. Shakura, Astron. Zh. 49, 921 (1972) [Sov. Astron.
16, 756 (1972)].

8. N. I. Shakura and R. A. Sunyaev, Astron. Astrophys.
24, 337 (1973).

9. G. S. Bisnovatyi-Kogan and S. I. Blinnikov, Astron.
Astrophys. 59, 111 (1977).

10. T. Okuda and M. Fujita, Publ. Astron. Soc. Jpn. 52,
L5 (2000).

11. G. I. Barenblatt, Similarity, Self-Similarity,
and Intermediate Asymptotics (Gidrometeoizdat,
Leningrad, 1978, 1982; Consultants Bureau, New
York, 1979).

12. W. Brinkman and N. Kawai, Astron. Astrophys. 363,
640 (2000).

13. H. C. Spruit, The Neutron Star—Black Hole Connection,
Ed. by C. Kouveliotou, J. Ventura, and
Ed van den Heuvel (Kluwer Acad., 2001), p. 141.

14. I. Yi, Astrophysical Disks, Ed. by J. A. Sellwood and
J. Goodman; Astron. Soc. Pac. Conf. Ser. 160, 279
(1999).

15. G. S. Bisnovatyi-Kogan, Discrete Dyn. Nature Soc.
6, 247 (2001).

16. R. D. Blandford and D. G. Payne, Mon. Not. R. Astron.
Soc. 199, 883 (1982).

17. R. E. Pudritz and C. A. Norman, Astrophys. J. 274,
677 (1983).

18. R. E. Pudritz and C. A. Norman, Astrophys. J. 301,
571 (1986).

19. G. Pelleter and R. E. Pudritz, Astrophys. J. 394, 117
(1992).

20. G. S. Bisnovatyi-Kogan and A. A. Ruzmaikin, Astrophys.
Space Sci. 28, 45 (1974).

21. J. Ferreira and G. Pelletier, Astron. Astrophys. 295,
807 (1995).

22. F. Casse and J. Ferreira, Astron. Astrophys. 353, 1115
(2000).

23. M. M. Romanova, G. V. Ustyugova, V. M. Chechetkin,
and R. V. E. Lovelace, Astrophys. J. 500, 703
(1998).

24. I. F. Mirabel, V. Dhawan, S. Chaty, et al., Astron.
Astrophys. 330, L9 (1998).

25. P. Murdin, D. H. Clark, and P. G. Martin, Mon. Not.
R. Astron. Soc. 193, 135 (1980).

26. S.V. Bogovalov and K. Tsinganos, Astron. Astrophys.
356, 989 (2000).

27. J. Contopoulos, Astrophys. J. 460, 185 (1996).

28. N. Vlahakis and K. Tsinganos, Mon. Not. R. Astron.
Soc. 307, 279 (1999).

29. Z. Y. Li, Astrophys. J. 444, 848 (1995).

30. V. S. Beskin, Usp. Fiz. Nauk 169, 689 (1997).

31. M. I. Pudovkin and V. S. Semenov,Ann. Geophys. 33,
429 (1977).

32. V. S. Semenov, in Geomagnetic Investigations
(Sov. Radio, Moscow, 1979) Vol. 24, p. 32 [in Russian].

33. E. Parker, Interplanetary Dynamical Processes
(Interscience, New York, 1963), p. 51.

34. S. V. Bogovalov, Mon. Not. R. Astron. Soc. 270, 721
(1994).

35. S. V. Bogovalov, Astron. Astrophys. 323, 634 (1997).

36. K. Tsinganos, C. Sauty, G. Surlantzis, et al., Mon.
Not. R. Astron. Soc. 283, 811 (1996).

37. N. Vlahakis, K. Tsinganos, C. Sauty, and E. Trussoni,
Astrophys. J. 545, 758 (2000).

38. T. Lery, R. Henriksen, and J. D. Fiege, Astron. Astrophys.
350, 254 (1999).

\newpage

\begin{figure}[h]
\centering{
\includegraphics[width=0.7\textwidth,angle=0,clip=]{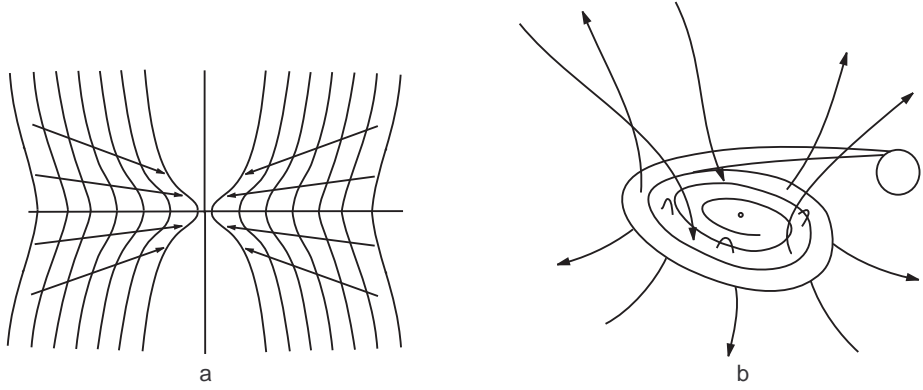}
\caption{\small Role of the magnetic field in different formulations of the problem.
 Left: plasma diffuses across the magnetic-field lines.
Right: plasma falls onto the gravitating center along the magnetic field.
The wind flowing from the disk extends some field lines
to infinity and carries away some fraction of the angular momentum of the
infalling material.}
\label{fig1}
}
\end{figure}

\vspace{1cm}

\begin{figure}[h]
\centering{
\includegraphics[width=0.5\textwidth,angle=0,clip=]{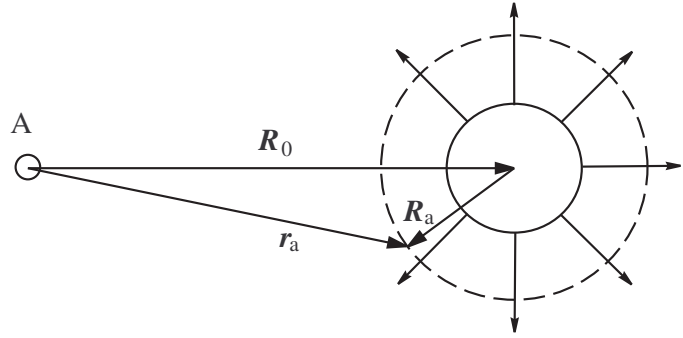}
\caption{\small Loss of angular momentum via a magnetized wind.
 The gravitating center is at the point A.}
\label{fig2}
}
\end{figure}

\vspace{1cm}

\begin{figure}[h]
\centering{
\includegraphics[width=0.7\textwidth,angle=0,clip=]{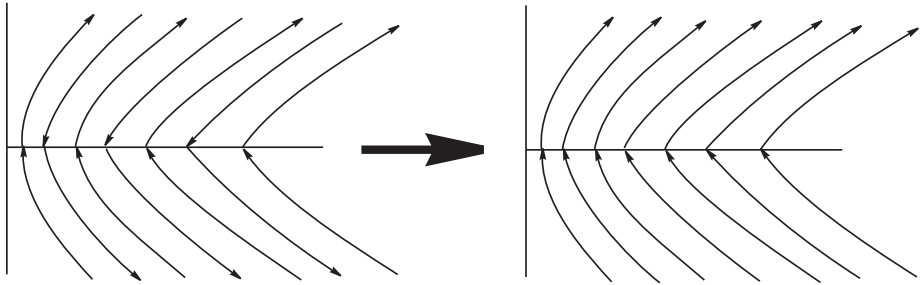}
\caption{\small Independence of the plasma dynamics of the direction
of the magnetic-field lines in an ideal MHD approximation. The
left panel shows a realistic structure for the flux of the poloidal
magnetic field; the total flux of the poloidal magnetic field is zero.
In the right panel, the direction of the field lines has changed,
while the plasma dynamics remain unaltered.}
\label{fig3}
}
\end{figure}


\begin{figure}[h]
\centering{
\includegraphics[width=0.4\textwidth,angle=0,clip=]{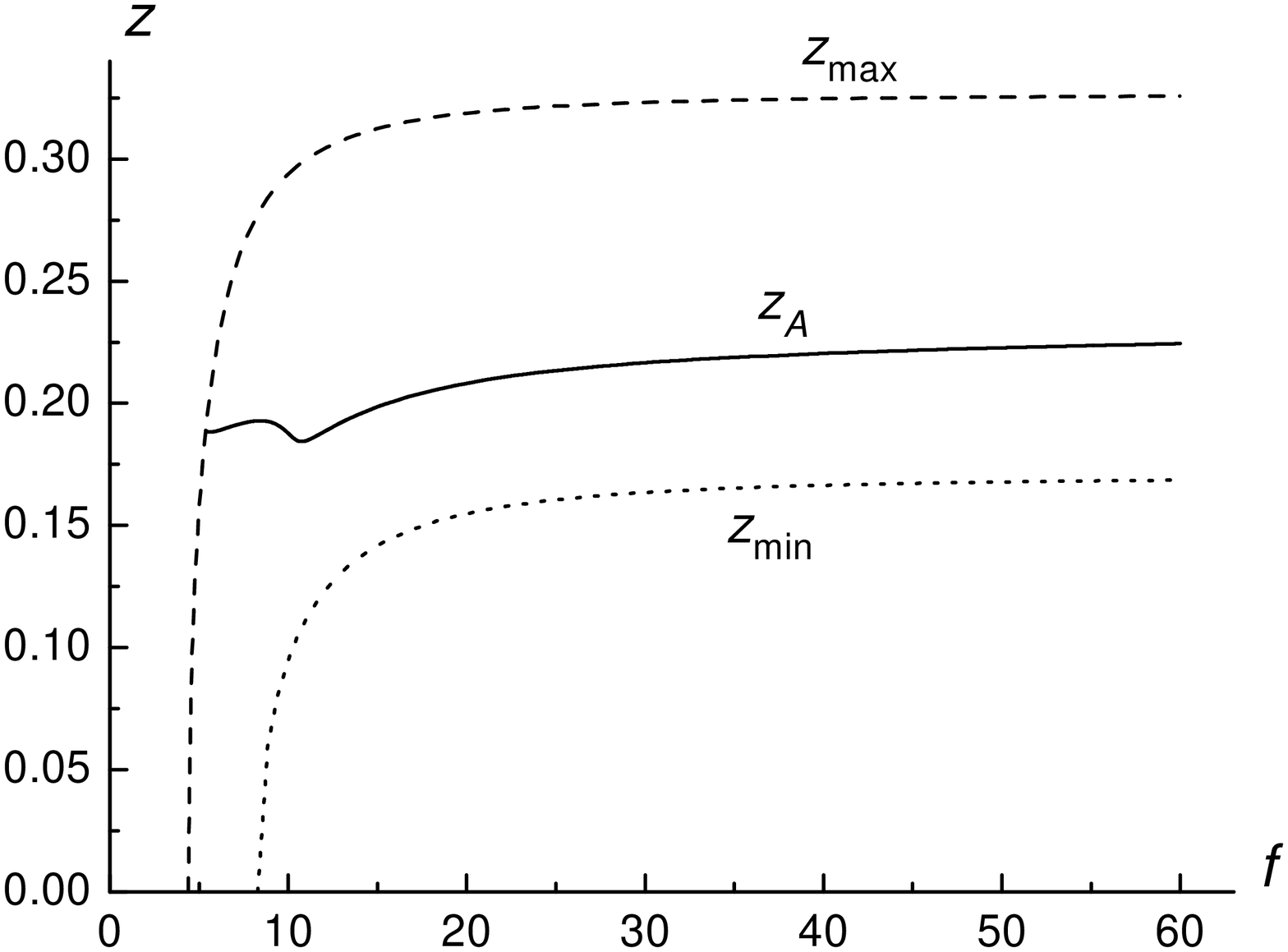}
\caption{\small Values $z_A$ of $\t z(f)$ for which a field line emerging
from the Alfven point crosses the equator when $\t r=1$.
$z_{\,\max}$ and $z_{\,\min}$, which appear in (95), are also indicated.}
\label{fig4}
}
\end{figure}

\vspace{1cm}

\begin{figure}[h]
\centering{
\includegraphics[width=0.4\textwidth,angle=0,clip=]{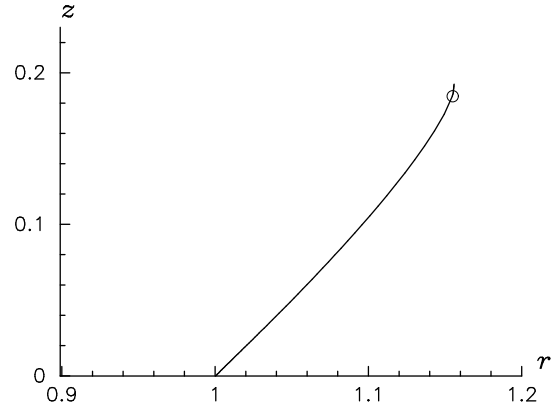}
\caption{\small A line of the poloidal field for $f=6$. The circle
marks the position of the Alfven singularity.}
\label{fig5}
}
\end{figure}

\vspace{1cm}

\begin{figure}[h]
\centering{
\includegraphics[width=0.4\textwidth,angle=0,clip=]{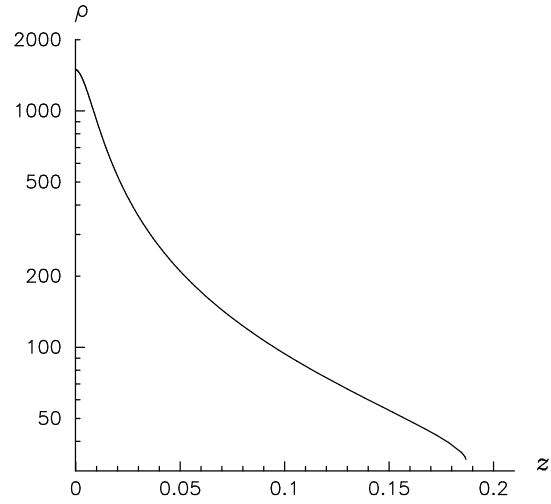}
\caption{\small $z$ dependence of the plasma density on the field line
shown in Fig. 5.}
\label{fig6}
}
\end{figure}

\end{document}